\documentclass{emulateapj}

\begin{document}

\shortauthors{Luhman et al.}
\shorttitle{Young Substellar Companion}

\title{Discovery of a Young Substellar Companion in Chamaeleon}

\author{K. L. Luhman\altaffilmark{1},
J. C. Wilson\altaffilmark{2},
W. Brandner\altaffilmark{3},
M. F. Skrutskie\altaffilmark{2},
M. J. Nelson\altaffilmark{2},
J. D. Smith\altaffilmark{4},
D. E. Peterson\altaffilmark{2},
M. C. Cushing\altaffilmark{4},
and E. Young\altaffilmark{4}}

\altaffiltext{1}{Department of Astronomy and Astrophysics,
The Pennsylvania State University, University Park, PA 16802;
kluhman@astro.psu.edu.}

\altaffiltext{2}{
Department of Astronomy, The University of Virginia, Charlottesville, 
VA 22903.}

\altaffiltext{3}{
Max-Planck-Institut f\"ur Astronomie, K\"onigstuhl 17, D-69117 Heidelberg, 
Germany.}

\altaffiltext{4}{
Steward Observatory, The University of Arizona, Tucson, AZ 85721.}

\begin{abstract}

During an imaging survey of the Chamaeleon~I star-forming region with
the Advanced Camera for Surveys aboard the {\it Hubble Space Telescope},
we have discovered a candidate substellar companion to the young low-mass 
star CHXR~73 ($\tau\sim2$~Myr, $M\sim0.35$~$M_\odot$).  We measure a 
projected separation of $1\farcs30\pm0\farcs03$ for the companion, 
CHXR~73~B, which corresponds to 210~AU at the distance of the cluster.
A near-infrared spectrum of this source obtained with the
Cornell Massachusetts Slit Spectrograph at the Magellan~II telescope
exhibits strong steam absorption that confirms its late-type nature
($\gtrsim$M9.5). In addition, the gravity-sensitive shapes of the 
$H$- and $K$-band continua demonstrate that CHXR~73~B is a young, 
pre-main-sequence object rather than a field star. The probability that 
CHXR~73~A and B are unrelated members of Chamaeleon~I is $\sim0.001$.
We estimate the masses of CHXR~73~B and other known substellar companions
in young clusters with a method that is consistent with the dynamical 
measurements of the eclipsing binary brown dwarf 2M~0535-0546, which 
consists of a comparison of the bolometric luminosities of the companions 
to the values predicted by the evolutionary models of Chabrier \& 
coworkers and Burrows \& coworkers. We arrive at mass estimates of 
0.003-0.004, $0.024\pm0.012$, $0.011^{+0.01}_{-0.003}$, and 
$0.012^{+0.008}_{-0.005}$~$M_{\odot}$ for 2M~1207-3932~B, GQ~Lup~B, 
DH~Tau~B, and CHXR~73~B, respectively.  Thus, DH~Tau~B and CHXR~73~B 
appear to be the least massive companions to stars outside the solar 
system that have been detected in direct images, and may have masses 
that are within the range observed for extrasolar planetary companions 
($M\lesssim0.015$~$M_{\odot}$). However, because these two objects 
(as well as 2M~1207-3932~B) probably did not form within circumstellar 
disks around their primaries, we suggest that they should be viewed as 
brown dwarf companions rather than planets.

\end{abstract}


\section{Introduction}
\label{sec:intro}

Since the first discovery of substellar companions outside of the solar system
little more than a decade ago \citep{wol92,mq95,mb96,opp95}, 
a great deal of work has focused on understanding the origin of these 
bodies. The three most widely explored theories for the formation of substellar 
companions involve the accretion of gas by rocky cores in circumstellar disks
\citep{pol96}, the collapse of gravitationally unstable areas of disks
\citep{bos98,bos06}, and the fragmentation of molecular cloud cores 
\citep{lod05}. Testing the validity of these models requires 
observations of companions across a large range 
of companion mass, primary mass, orbital separation, and age. 
Searches for substellar companions at young ages are particularly valuable 
because they can directly constrain the formation timescale and early
dynamical evolution of these objects.
In addition, substellar objects are brightest when they are
young, and thus can be detected at very low masses with direct imaging. 
For these reasons, young stars and brown dwarfs in the solar neighborhood
\citep[$\tau\sim30$-400~Myr,][]{reb98,mh04,mz04,low05},
open clusters and associations
\citep[$\tau\sim10$-100~Myr,][]{mar98,mar00,mar03,low00,neu00a,gue01,cha03,cha05b},
and star-forming regions and OB associations
\citep[$\tau\lesssim5$~Myr,][]{duc99,bra00,neu02,luh05wfpc,kra05,kra06}
have been popular targets for companion searches with high-resolution imaging.
The most reliable substellar companions discovered to date at ages of
$\tau<10$~Myr consist of GG~Tau~Bb \citep{whi99}, TWA~5~B \citep{low99,neu00b},
2M~1101-7732~B \citep{luh04bin}, 2M~1207-3932~B \citep{cha04,cha05a},
DH~Tau~B \citep{ito05}, and GQ~Lup~B \citep{neu05}.

Adding to the small but quickly growing list of known young substellar 
companions, we report the discovery of a companion to the low-mass star
CHXR~73 \citep[M3.5, $M\sim0.35$~$M_\odot$,][]{luh04cha} in the Chamaeleon~I 
star-forming region ($\tau\sim2$~Myr), which was found serendipitously during 
a survey for free-floating brown dwarfs with the {\it Hubble Space Telescope}
($HST$).  In this paper, we present our optical images from $HST$ and 
near-infrared (IR) imaging and spectroscopy from the Magellan telescopes
for CHXR~73~B (\S~\ref{sec:obs}) and we use these data to assess its youth 
and membership in Chamaeleon~I (\S~\ref{sec:mem}). We then examine the
evidence that CHXR~73~B is a companion (\S~\ref{sec:bin}), 
estimate the mass of CHXR~73~B and other young substellar companions in a 
uniform manner (\S~\ref{sec:mass}), and discuss the origin of these companions 
and whether they should be referred to as planets or brown dwarfs
(\S~\ref{sec:disc}).

\section{Observations}
\label{sec:obs}

\subsection{Photometry}
\label{sec:phot}

\begin{figure}
\plotone{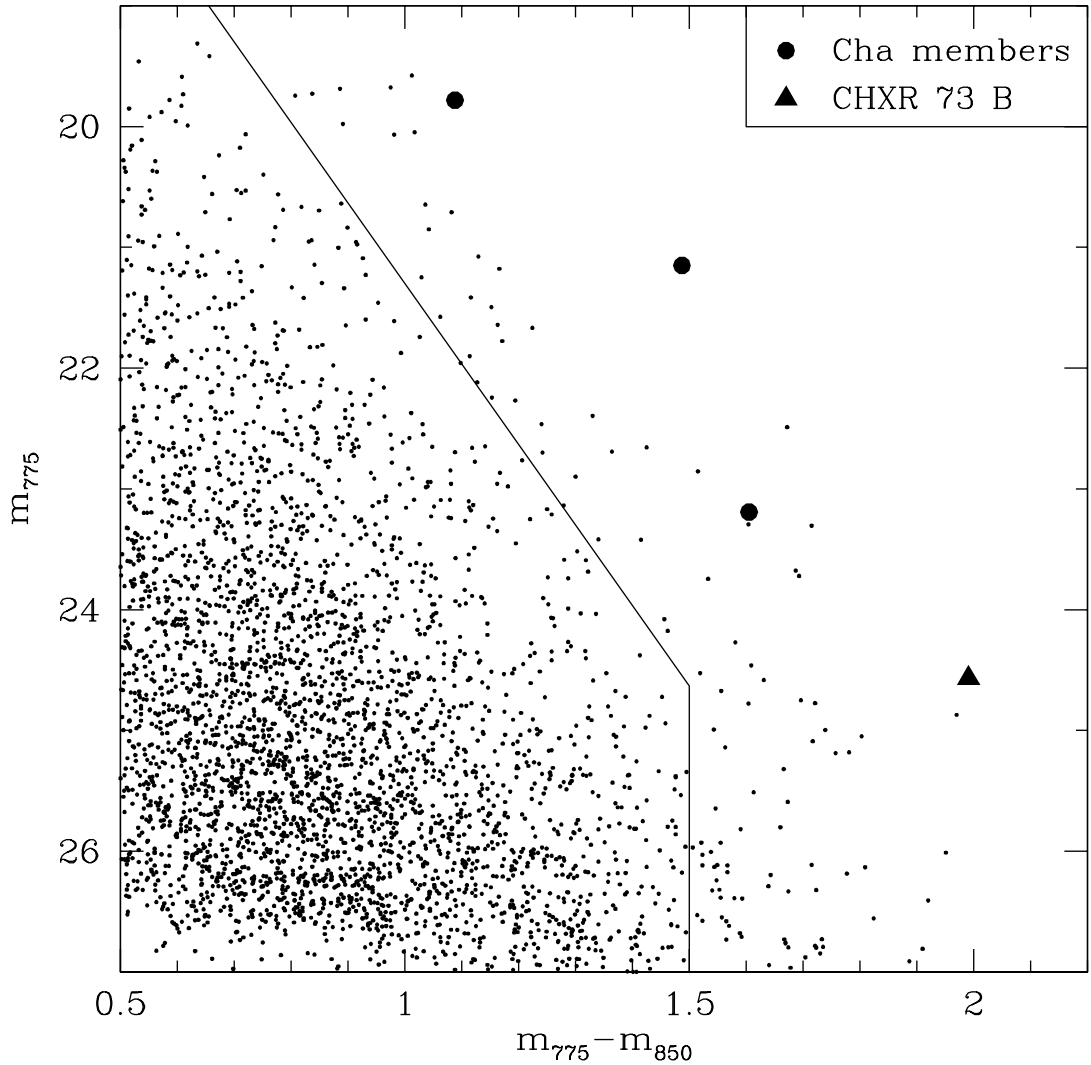}
\caption{ 
Color-magnitude diagram for unsaturated stars in ACS images of a
$13\farcm3\times16\farcm7$ area in the southern cluster of the 
Chamaeleon~I star-forming region. 
We indicate the unsaturated previously known cluster members in these data 
({\it large points}) and the candidate companion to CHXR~73 shown in 
Figure~\ref{fig:images} ({\it triangle}).
The solid boundary was designed to follow the lower envelope of the sequence
of known members.
}
\label{fig:iz}
\end{figure}

\begin{figure}
\plotone{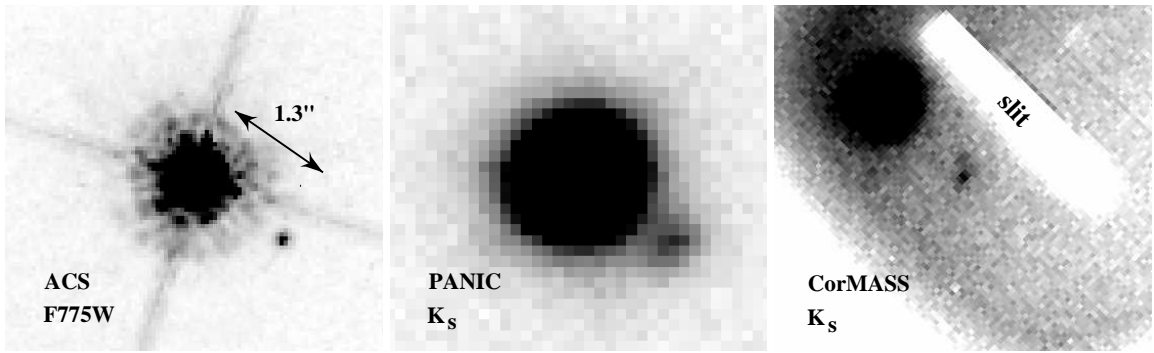}
\caption{ 
Images of CHXR~73~A and B obtained with ACS on {\it HST} (F775W), 
PANIC on Magellan~I ($K_s$), and CorMASS on Magellan~II ($K_s$). 
After the CorMASS exposure, the spectrograph's slit was centered on 
CHXR~73~B for spectroscopic observations (Figure~\ref{fig:ir}).
These images exhibit FWHM$=0\farcs1$, $0\farcs5$, and $0\farcs2$, respectively,
and have dimensions of $4\arcsec\times4\arcsec$. North is up and east is left
in these images.
}
\label{fig:images}
\end{figure}

As a part of a survey for young free-floating brown dwarfs, 
\citet{luh05cha} used the Advanced Camera for Surveys (ACS) aboard {\it HST}
to image a $13\farcm3\times16\farcm7$ area centered on the southern cluster
in the Chamaeleon~I star-forming region through the F775W and F850LP filters
(0.775 and 0.85~\micron).
In Figure~\ref{fig:iz}, we plot the color-magnitude diagram produced by those
data. Most members of the cluster are saturated in the ACS images
and thus do not appear in Figure~\ref{fig:iz}, including most of the
known young brown dwarfs within the survey field.
The faintest known member in the diagram is Cha~110913-773444, which was
discovered with these data by \citet{luh05cha}.
As done in a similar color-magnitude diagram constructed from $HST$ data
for the young cluster IC~348 \citep{luh05wfpc}, we define a boundary below 
the lower envelope of the sequence of known members of Chamaeleon~I 
to separate candidate cluster members and probable field stars. 
The reddest brown dwarf candidate in Figure~\ref{fig:iz} has a separation of 
only $1\farcs3$ from a known cluster member, CHXR~73, making it a promising 
substellar companion. We refer to this object as CHXR~73~B hereafter in this 
paper. The ACS images containing CHXR~73~B were obtained on 2005 February 10.
A $4\arcsec\times4\arcsec$ area of the F775W image encompassing 
CHXR~73~A and B is shown in Figure~\ref{fig:images}.
The ACS photometric measurements for this object are $m_{775}=24.57\pm0.03$ and
$m_{850}=22.58\pm0.03$.
CHXR~73~B has a separation of $1\farcs30\pm0\farcs03$ and a position angle of 
$234.9\pm1\arcdeg$ from the primary. The errors in these astrometric
measurements are dominated by the error in the position of CHXR~73~A, which 
is saturated. 
Using a distance modulus of 6.05 for Chamaeleon~I \citep{whi97,wic98,ber99}, 
this projected separation corresponds to 210~AU.

After discovering CHXR~73~B in the ACS images, we obtained photometry for it
at near-IR wavelengths with Persson's Auxiliary Nasmyth Infrared Camera 
(PANIC) on the Magellan~I telescope on the night of 2005 April 30.
The instrument contained one $1024\times1024$ HgCdTe Hawaii array with a 
plate scale of $0\farcs126$~pixel$^{-1}$. 
We obtained 10, 10, and 20 dithered images in the filters $J$, $H$, and $K_s$, 
respectively, with exposure times of 3~sec.
After the data at each filter were dark subtracted, flat fielded, and
combined, the final images exhibited FWHM$=0\farcs5$ for point sources. The 
combined image of CHXR~73~A and B in $K_s$ is shown in Figure~\ref{fig:images}.
We flux calibrated the images with photometry from \citet{luh05cha} 
for unsaturated stars in these images and measured photometry of 
$J=17.9\pm0.3$, $H=16.5\pm0.3$, and $K_s=15.5\pm0.25$ for CHXR~73~B, 
corresponding to $\Delta J=5.2$, $\Delta H=5.2$, and $\Delta K_s=4.7$ 
relative to the primary.

\subsection{Spectroscopy}

We obtained near-IR spectra of CHXR~73~B with the Cornell Massachusetts 
Slit Spectrograph \citep[CorMASS,][]{wil01} on the Magellan~II telescope
during the nights of 2005 April 30 and May 1. This instrument provides 
simultaneous wavelength coverage from 0.8-2.5~\micron\ and a resolution of 
$R\sim300$.  While mounted on the Magellan~II telescope, the slit of CorMASS 
subtended an area of $0\farcs4\times3\arcsec$ on the sky. 
To observe CHXR~73~B, we adjusted the position angle of the slit to align it
with the axis connecting CHXR~73~A and B. Figure~\ref{fig:images}
shows an image of the pair taken in the slit-viewing mode of CorMASS 
with a $K_s$ filter and an exposure time of 5~sec prior to starting
spectroscopic observations. We then selected an exposure time
of 3~min and obtained two exposures with CHXR~73~B centered on the slit,
one exposure at a position several arcseconds from CHXR~73~B, and
repeated this cycle two more times, resulting in total exposure times of
18 and 6~min on CHXR~73~B and the sky, respectively. This sequence was
performed on both nights. A nearby A0~V star (HD~98671) was also observed 
for the correction of telluric absorption. 
The spectra were reduced with a modified version of the Spextool package
\citep{cus04} and were corrected for telluric absorption \citep{vac03}.
The spectra from the two nights agreed well and thus were combined.
For comparison, we also obtained spectra of the field dwarfs Gl~406, 
LHS~2065, and Kelu~1.

\section{Analysis}
\label{sec:analysis}

\subsection{Evidence of Membership in Chamaeleon~I}
\label{sec:mem}

\begin{figure}
\plotone{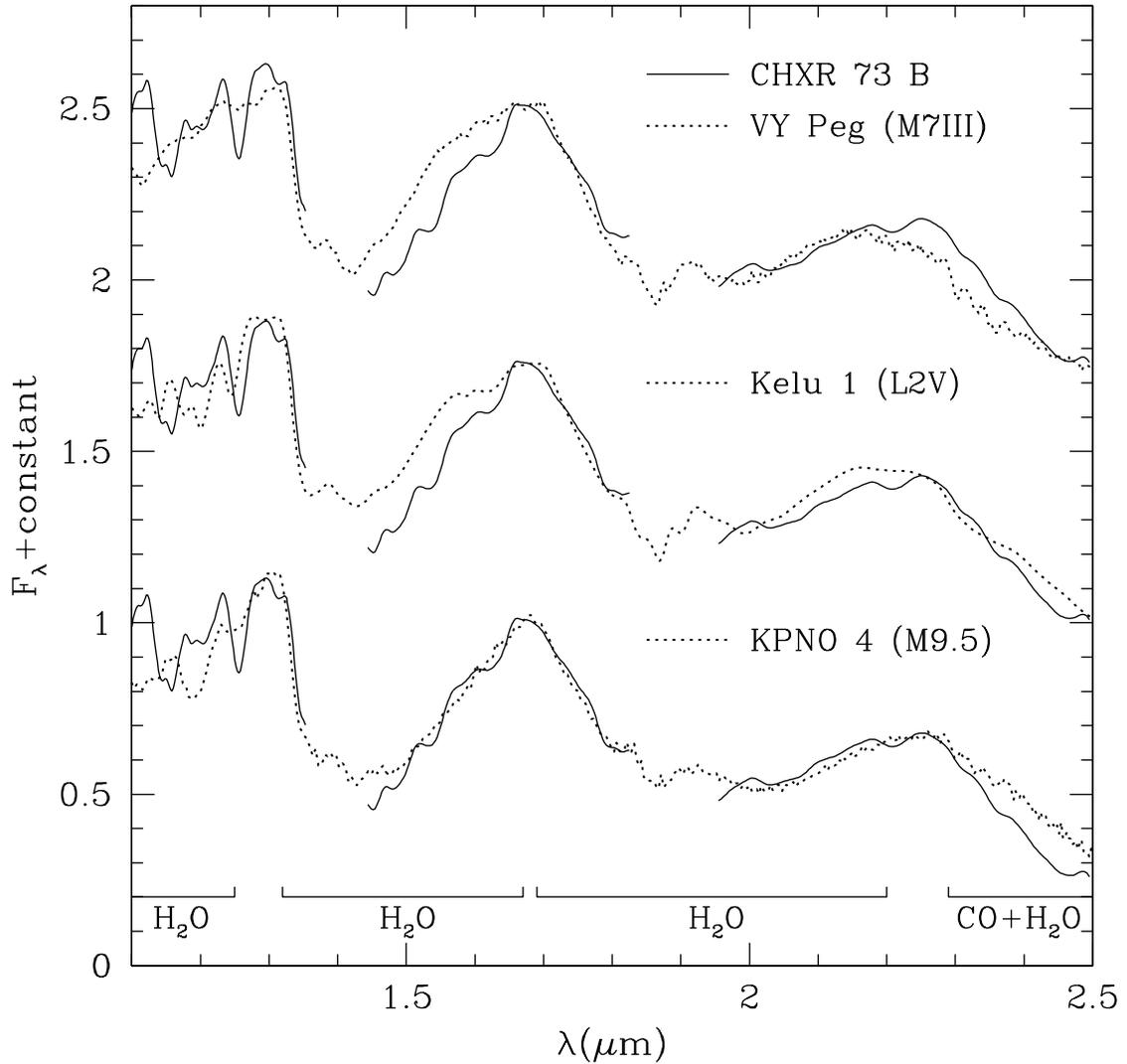}
\caption{
CorMASS infrared spectrum of the candidate substellar companion CHXR~73~B 
compared to spectra of the field M giant VY~Peg, the field L dwarf Kelu~1, 
and the young Taurus member KPNO~4 ($\tau\sim1$~Myr). 
CHXR~73~B exhibits the triangular $H$- and $K$-band continua that are
indicative of young late-type objects rather than the broad plateaus
found in the field objects, confirming its youth and membership in the 
Chamaeleon~I star-forming region. 
The spectrum of CHXR~73~B has been dereddened by $A_J=2.1$ to match the
slope of KPNO~4, which has $A_J\sim0$.
To facilitate comparison to CHXR~73~B, the spectra of VY~Peg and Kelu~1
were then reddened by $A_J=0.4$ to match the slope of this dereddened 
spectrum of CHXR~73~B.
Because the observed spectrum of CHXR~73~B is very red, 
the signal-to-noise ratio is lower at shorter wavelengths, and thus
the continuum structure at $<$1.3~\micron\ is noise rather than 
photospheric absorption features. 
The spectra are displayed at a resolution of $R=100$ and are normalized at 
1.68~\micron.}
\label{fig:ir}
\end{figure}

We now use the spectroscopy from the previous section to determine if 
CHXR~73~B is a member of the Chamaeleon~I star-forming region 
rather than a field star or a galaxy. 
To classify the spectrum, we compared it to data for field dwarfs and
giants and known members of star-forming regions. For the field dwarfs, 
we used the CorMASS spectra of Gl~406, LHS~2065, and Kelu~1. We employed as 
additional standards our previous spectra of late-type dwarfs, giants, and 
pre-main-sequence objects obtained with SpeX at the NASA Infrared Telescope 
Facility \citep{luh05wfpc,luh04ots,luh06tau}.
For the two standards that appear in both the CorMASS and SpeX samples, 
Gl~406 and LHS~2065, the spectra from the two instruments agree well,
which suggests that SpeX data are suitable for 
classifying the CorMASS spectrum of CHXR~73~B. 
In Figure~\ref{fig:ir}, we compare the spectrum of CHXR~73~B to a standard
in each of the three luminosity classes, namely the field dwarf 
Kelu~1 \citep[L2V,][]{kir99}, the field giant VY~Peg \citep[M7III,][]{kir97}, 
and the young object KPNO~4 \citep[M9.5,][]{bri02}.
To facilitate the comparison of the spectra, the spectral slopes of the former
three objects have been aligned with that of KPNO~4 
\citep[$A_J\sim0$,][]{bri02} by reddening or dereddening the spectra
according to the extinction law of \citet{rl85}. 
To match KPNO~4, the spectrum of CHXR~73~B must be dereddened by $A_J\sim2.1$. 

Like the late-type standards in Figure~\ref{fig:ir}, CHXR~73~B exhibits 
strong H$_2$O absorption bands, demonstrating that it is a cool object rather 
than an early-type field star or an extragalactic source. 
To determine the luminosity class of CHXR~73~B, we can examine
the shapes of the $H$- and $K$-band continua, which are sensitive to 
surface gravity \citep{luc01,luh04ots,kir06}.
The continua of CHXR~73~B have the same triangular shape as the young
brown dwarf KPNO~4 rather than the broad plateaus that characterize
the dwarf and the giant. Based on this comparison, we conclude that
CHXR~73~B is a young member of Chamaeleon~I rather than a field star.
The presence of significant extinction toward CHXR~73~B is independent evidence
that it is not a foreground object. In addition, with an extinction-corrected
magnitude of $K=14.7$, CHXR~73~B is too bright to be a background field dwarf
given that field dwarfs later than M9V have $M_K>10$ \citep{dah02,gol04}
and Chamaeleon~I has a distance modulus of 6.05.

The strengths of the H$_2$O absorption bands in the spectrum of CHXR~73~B
are equal to or slightly greater than those of KPNO~4, which has an optical 
spectral type of M9.5. Because the variation of H$_2$O absorption with
optical spectral type is unknown for young objects later than M9, we can place
only a limit of $\geq$M9.5 on the spectral type. If CHXR~73~B is later than
KPNO~4, then it probably has a redder intrinsic spectrum than KPNO~4, 
in which case CHXR~73~B would have an extinction lower than the value of
$A_J=2.1$ implied by the comparison to KPNO~4. For instance, the difference
in spectral slopes between LHS~2065 (M9V) and Kelu~1 (L2V) is equivalent to
$A_J\sim0.15$. Therefore, we adopt an extinction of $A_J=2\pm0.3$ for CHXR~73~B.

\subsection{Evidence of Binarity}
\label{sec:bin}

In the previous section, we demonstrated that CHXR~73~B is a member of 
Chamaeleon~I. We now examine the evidence that it is a companion to CHXR~73~A.
The $13\farcm3\times16\farcm7$ area imaged in our ACS survey encompasses 
39 previously known cluster members and six new low-mass members
($m_{775}>22$, $M\lesssim0.015$~$M_\odot$). The latter sources consist of
Cha~110913-773444, CHXR~73~B, and four unpublished objects. 
The probability of any of these six sources having a projected separation
less than $1\farcs3$ from any of the 39 higher mass members in the ACS
survey area is $\sim0.001$. Because of the low value of this probability, 
we conclude that CHXR~73~B is a companion to CHXR~73~A rather than 
an unrelated cluster member. 
The available evidence of binarity for each of the other young substellar 
companions discussed in this work is also based on statistical analysis of 
this kind because the published proper motion measurements for those 
objects are not precise enough to distinguish between true binaries and 
comoving unbound cluster members seen in projection near each other. 
For instance, consider a pair of unrelated cluster members with    
two-dimensional velocities differing by 0.5~km~s$^{-1}$, which is comparable
to velocity dispersions in star-forming regions. During the five years 
between the two epochs of images of GQ~Lup~B from \citet{neu05}, the
separation of this pair would change by $0\farcs003$, which is below the 
precision of the astrometry from \citet{neu05}.
Finally, we note that the extinction of CHXR~73~B ($A_J=2$) is similar to that
of CHXR~73~A \citep[$A_J=1.8$,][]{luh04cha}, which further supports the
binarity of the pair.

\subsection{Mass}
\label{sec:mass}

To estimate the mass of CHXR~73~B, we compare its bolometric luminosity to
the values predicted for young brown dwarfs by the theoretical evolutionary 
models of \citet{cha00} and \citet{bur97}. 
We test the validity of this mass estimate by performing the same analysis 
with the eclipsing binary brown dwarfs 2M~0535-0546~A and B,
whose masses have been measured dynamically \citep{sta06}.
To place CHXR~73~B in the context of other young low-mass companions,
we also include in this exercise 2M~1207-3932~B, DH~Tau~B, and GQ~Lup~B.

We compute the bolometric luminosities of 2M~0535-0546~A and B by 
combining their temperature ratio and radii \citep{sta06} with 
an optical spectral type of M6.75 for A+B (K. Luhman, in preparation) and
a temperature scale for young objects \citep{luh03}. 
For each of the remaining companions, we estimate the luminosity
from a combination of the observed $K$ magnitude, extinction, distance, 
$K$-band bolometric correction \citep{gol04}, and an absolute bolometric 
magnitude for the Sun of $M_{\rm bol \odot}=4.75$.
For CHXR~73~B, we adopt $K=15.5\pm0.25$ (\S~\ref{sec:phot}), $A_J=2\pm0.3$
(\S~\ref{sec:mem}), a distance modulus of $6.05\pm0.13$, and BC$_K=3.2\pm0.18$
for an assumed spectral type of M9-L1. 
For 2M~1207-3932~B, we adopt $M_K=13.30\pm0.29$ from \citet{mam05}, which is 
based on his distance estimate and the photometry reported by \citet{cha04}.
At a given optical spectral type, near-IR steam absorption bands are stronger 
in young objects than in field dwarfs \citep{luc01,luh04ots}, as
illustrated in Figure~\ref{fig:ir}.  As a result, 
because \citet{cha04} classified 2M~1207-3932~B by comparing the strength of
its steam absorption to that of field dwarfs, their spectral classification 
of L5-L9.5 is probably systematically too late. Therefore, we assume an
earlier range of possible types, L0-L7, and adopt the 
corresponding bolometric correction of BC$_K=3.23\pm0.16$.
For GQ~Lup~B, we adopt the distance of $150\pm20$~pc measured for Lupus~I 
\citep{hug93,cra00,fra02}, $K_s=13.1\pm0.1$ \citep{neu05}, and the extinction
of $A_V=0.4\pm0.2$ measured for the primary \citep{bat01}.
As with 2M~1207-3932~B, the IR classification of GQ~Lup~B from \citet{neu05}
was performed through a comparison of its steam absorption to that of dwarf
standards, making the resulting spectral type too late. 
Because GQ~Lup~B is similar to known late-type Taurus members in terms of $M_K$,
it probably has a similar spectral type. Therefore, we adopt the bolometric
correction that applies to M8-L0, BC$_K=3.15\pm0.18$.
For DH~Tau~B, we adopt $K_s=14.19\pm0.02$ and $A_J=0.3\pm0.3$ 
\citep{ito05} and a distance modulus of $5.76\pm0.2$ \citep{wic98}.
Because $M_K$ for DH~Tau~B is similar to that of CHXR~73~B, the value of 
BC$_K$ for CHXR~73~B is also applied to DH~Tau~B.
Through the above calculations, we arrive at log~$L_{\rm bol}=-2.85\pm0.14$,
$-1.58\pm0.07$, $-1.73\pm0.07$, $-4.71\pm0.13$, $-2.71\pm0.12$, 
and $-2.23\pm0.14$ for CHXR~73~B, 2M~0535-0546~A and B, 2M~1207-3932~B, 
DH~Tau~B, and GQ~Lup~B, respectively.

We must adopt an age for each companion to convert its luminosity to a mass 
with the evolutionary models.
On the Hertzsprung-Russell, the median age of the known members of 
Chamaeleon~I is 2~Myr \citep{luh04cha} and the position of CHXR~73~A also 
implies an age of 2~Myr. Therefore, we adopt this value for CHXR~73~B.
The star-forming regions containing DH~Tau~B, GQ~Lup~B, and 
2M~0535-0546~A and B appear to be slightly younger with ages closer
to $\sim1$~Myr \citep{hil97,luh03}. For reasons described by \citet{luh05cha}, 
we adopt conservative lower and upper age limits 
of 0.5 to 10~Myr for the above companions in star-forming clusters.
Meanwhile, for 2M~1207-3932~B we assume an age of $8^{+4}_{-3}$~Myr as done
by \citet{cha04}.

In Figure~\ref{fig:lbol}, we plot the positions of CHXR~73~B, 2M~0535-0546~A 
and B, 2M~1207-3932~B, DH~Tau~B, and GQ~Lup~B with the luminosities predicted
as a function of age for masses from 0.005-0.06~$M_{\odot}$ by \citet{bur97}
and \citet{cha00}. For an age of 1~Myr, these models imply masses of 
0.055-0.065 and 0.05-0.06~$M_{\odot}$ for 2M~0535-0546~A, which agrees
with the dynamical measurement of 0.05-0.059~$M_{\odot}$ \citep{sta06}.
Meanwhile, the model estimates of 0.045-0.055 and 0.04-0.05~$M_{\odot}$ for 
2M~0535-0546~B are higher than the dynamical mass of 0.031-0.037~$M_{\odot}$.
The models of \citet{bur97} are consistent with the dynamical masses of 
both components if they are younger than 1~Myr. 
These data for 2M~0535-0546 indicate that the evolutionary models
produce reasonably accurate mass estimates for young brown dwarfs when 
they are derived from bolometric luminosities, 
confirming the results of other recent observational tests
\citep{luh05abdor,luh06}\footnote{However, the relative masses implied by
the temperature ratio of 2M~0535-0546~A and B are
inconsistent with the dynamical measurements, which may indicate that
the components are not coeval or that they are subject to a physical 
process that is not adequately treated by the models \citep{sta06}.}.
Thus, our mass estimates for the other young
low-mass companions in Figure~\ref{fig:lbol} should not be wildly in error,
although the models are untested at the lowest masses and could have larger
errors in that regime. 

According to Figure~\ref{fig:lbol}, 2M~1207-3932~B has a mass of 
0.003-0.004~$M_{\odot}$. We used the same data and models for this object
as \citet{mam05} except for a slightly different bolometric correction,
and thus arrived at the same mass.
For GQ~Lup~B, the models imply a mass of $0.024\pm0.012$~$M_{\odot}$ if
it has an age less than $\sim7$~Myr, which is similar to the values 
estimated by \citet{neu05} and \citet{jan06} from the luminosity and the same 
sets of models. 
For DH~Tau~B and CHXR~73~B, we derive masses of $0.011^{+0.01}_{-0.003}$ and
$0.012^{+0.008}_{-0.005}$~$M_{\odot}$, respectively.
Finally, although we are focusing on the youngest substellar companions, it 
is useful to also consider AB~Pic~B \citep{cha05b}, which is probably the 
least massive companion directly imaged at ages of $\tau>10$~Myr. 
The combination of its $K$-band magnitude \citep{cha05b}, an appropriate 
bolometric correction \citep{gol04}, and the distance of the primary 
\citep{per97} produce log~$L_{\rm bol}=-3.7\pm0.1$, which corresponds to
0.013-0.014~$M_{\odot}$ for an age of $\tau=30$~Myr based on the
models in Figure~\ref{fig:lbol}. Thus, the mass of AB~Pic~B appears to be
similar to, or perhaps slightly greater than, those of DH~Tau~B and CHXR~73~B.

\section{Discussion}
\label{sec:disc}

\begin{figure}
\plotone{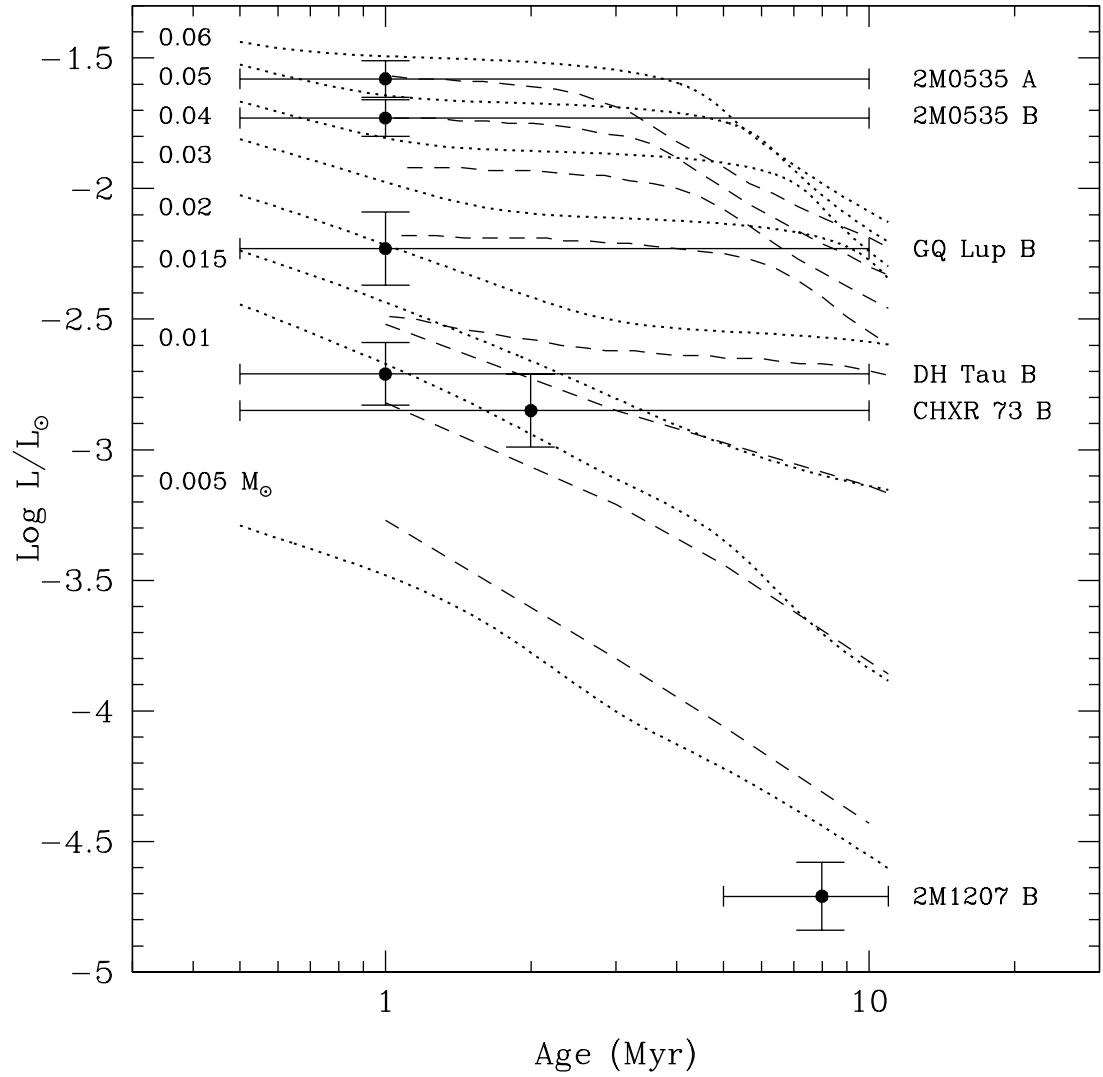}
\caption{ 
Luminosities of the young companions 2M~0535-0546~A and B, 
GQ~Lup~B, 2M~1207-3932~B, DH~Tau~B, and CHXR~73~B ({\it points}) compared 
to the luminosities as a function of age predicted by the theoretical 
evolutionary models of \citet{cha00} ({\it dashed lines}) 
and \citet{bur97} ({\it dotted lines}) for masses of 0.005 to 
0.06~$M_{\odot}$. CHXR~73~B has a mass of 
$0.012^{+0.008}_{-0.005}$~$M_{\odot}$ according to these models.
}
\label{fig:lbol}
\end{figure}

We have estimated masses for CHXR~73~B and other young low-mass companions
in a uniform manner, and with a method that produces masses that are 
consistent with dynamical measurements for the one known young eclipsing binary 
brown dwarf, which allows us to examine both the relative and absolute masses 
of these companions. \citet{neu05} reported that 
GQ~Lup~B could have a mass as low as 1~$M_{\rm Jup}$, possibly making it 
the first planetary-mass companion imaged around a star. 
However, GQ~Lup~B is only slightly less luminous than 2M~1101-7732~B 
\citep{luh04bin} and is significantly brighter than 
DH~Tau~B \citep{ito05}.
If 2M~1101-7732~B and 2M~1207-3932~A were placed on the diagram of luminosity 
versus age in Figure~\ref{fig:lbol}, they would have masses comparable to
that of GQ~Lup~B. 
For added perspective, we note that GQ~Lup~B has a higher luminosity 
than many of the known late-type members of nearby star-forming regions 
\citep[$>$M8,][]{bri02,luh03}. 
Thus, it appears that GQ~Lup~B was not the least massive companion imaged at 
the time of its discovery and our mass estimate of $0.024\pm0.012$~$M_{\odot}$ 
suggests that it is unlikely to be a planet by any definition of the term. 

DH~Tau~B and CHXR~73~B are the faintest known companions to young stars 
($\tau<10$~Myr). If our mass estimates of $0.011^{+0.01}_{-0.003}$ and
$0.012^{+0.008}_{-0.005}$~$M_{\odot}$ are accurate, then they (and possibly
AB~Pic~B) are the least massive companions directly imaged near stars outside 
the solar system.
In fact, these estimates are just below both the maximum mass observed for
extrasolar planetary companions \citep[$M\sim0.015$~$M_{\odot}$,][]{mar05} and
the deuterium burning limit \citep[$M\sim0.014$~$M_{\odot}$,][]{sau96,cha00}. 
If DH~Tau~B and CHXR~73~B have planetary masses, then are they planets?
We can pose the same question for 2M~1207-3932~B, which has an even lower
mass according to the models. In other words, is a planetary-mass companion 
always a planet?

The definition of a planet can be based on a companion's mass or on its
formation mechanism. Following \citet{cha06}, we advocate the latter definition
in which a planet is an object that forms in the disk around a star or a 
brown dwarf, while a substellar companion that forms in another manner 
(e.g., core fragmentation) is a brown dwarf.
We apply this definition of a planet to DH~Tau~B, CHXR~73~B, and 2M~1207-3932~B
by drawing on theoretical work on planet formation in disks.
The maximum separations at which giant planet formation can occur via both
core accretion \citep[$\lesssim10$~AU,][]{pol96,ina03} and
disk instability \cite[$\lesssim20$-100~AU,][]{bos03,bos06}
are much lower than the projected separations of 200 and 300~AU for CHXR~73~B 
and DH~Tau~B, respectively. In addition, these systems are too young 
($\tau\sim1-2$~Myr) for the production of planets through core accretion.
Thus, CHXR~73~B and DH~Tau~B probably did not form in disks 
around their primaries. Formation via disk instability cannot be ruled out 
for 2M~1207-3932~B \citep{lod05}, but it would require an unusually high
ratio of disk mass to brown dwarf mass for 2M~1207-3932~A when it was younger
\citep{kle03,sch06}. Meanwhile, it has been shown that cloud fragmentation 
can produce isolated objects that reach into the planetary mass regime. 
For instance, the least massive free-floating brown dwarf in Chamaeleon~I is 
one magnitude fainter than CHXR~73~B 
\citep[$M\sim0.08$~$M_{\odot}$,][]{luh05cha}, so it seems plausible that the 
same process produced CHXR~73~B, but in a binary system instead of in isolation.
Based on these various theoretical and observational considerations,
we conclude that CHXR~73~B and other known young substellar companions 
probably did not form in a planetary fashion in disks and instead
should be referred to as brown dwarfs. However, because the products of planet 
formation and core fragmentation may overlap in separation as well as in mass
(e.g., 5~$M_{\rm Jup}$ at 10~AU), future surveys may uncover companions 
that cannot be classified unambiguously as either planets or brown dwarf
companions.

\acknowledgements
We thank Alan Boss for helpful discussions.
K. L. was supported by grant NAG5-11627 from the NASA Long-Term Space
Astrophysics program and grant GO-10138 from the Space Telescope Science
Institute.
CorMASS was supported by a generous gift to the University of Virginia Astronomy
Department from the F.H. Levinson Fund of the Peninsula Community Foundation.

\clearpage

\begin{thebibliography}{}

\bibitem[Batalha et al.(2001)]{bat01}
Batalha, C., Lopes, D. F., \& Batalha, N. M. 2001, \apj, 548, 377

\bibitem[Bertout et al.(1999)]{ber99}
Bertout, C., Robichon, N., \& Arenou, F. 1999, \aap, 352, 574

\bibitem[Boss(1998)]{bos98}
Boss, A. 1998, \apj, 503, 923

\bibitem[Boss(2003)]{bos03}
Boss, A. 2003, \apj, 599, 577

\bibitem[Boss(2006)]{bos06}
Boss, A. 2006, \apj, 637, L137

\bibitem[Brandner et al.(2000)]{bra00}
Brandner, W., et al. 2000, \aj, 120, 950

\bibitem[Brice\~no et al.(2002)]{bri02}
Brice\~{n}o, C., Luhman, K. L., Hartmann, L., Stauffer, J. R., \& Kirkpatrick,
J. D. 2002, \apj, 580, 317

\bibitem[Burrows et al.(1997)]{bur97}
Burrows, A., et al. 1997, \apj, 491, 856

\bibitem[Chabrier et al.(2000)]{cha00}
Chabrier, G., Baraffe, I., Allard, F., \& Hauschildt, P. 2000, \apj, 542, L119

\bibitem[Chabrier et al.(2006)]{cha06}
Chabrier, G., Baraffe, I., Selsis, F., Barman, T. S., Hennebelle, P., \&
Alibert, Y. 2006, Protostars and Planets V, in press

\bibitem[Chauvin et al.(2003)]{cha03}
Chauvin, G., et al. 2003, \aap, 404, 157

\bibitem[Chauvin et al.(2004)]{cha04}
Chauvin, G., et al. 2004, \aap, 425, L29

\bibitem[Chauvin et al.(2005a)]{cha05a}
Chauvin, G., et al. 2005a, \aap, 438, L25

\bibitem[Chauvin et al.(2005b)]{cha05b}
Chauvin, G., et al. 2005b, \aap, 438, L29


\bibitem[Crawford(2000)]{cra00}
Crawford, I. A. 2000, \mnras, 317, 996

\bibitem[Cushing et al.(2004)]{cus04}
Cushing, M. C., Vacca, W. D., \& Rayner, J. T. 2004, \pasp, 116, 362

\bibitem[Dahn et al.(2002)]{dah02}
Dahn, C. C., et al. 2002, \aj, 124, 1170

\bibitem[Duch\^ene et al.(1999)]{duc99}
Duch\^ene, G., Bouvier, J., \& Simon, T. 1999, 343, 831

\bibitem[Franco(2002)]{fra02}
Franco, G. A. P. 2002, \mnras, 331, 474


\bibitem[Golimowski et al.(2004)]{gol04}
Golimowski, D. A., et al. 2004, \aj, 127, 3516

\bibitem[Guenther et al.(2001)]{gue01}
Guenther, E. W., Neuh\"{a}user, R., Hu\'{e}lamo, N., Brandner, W., \& Alves, J.
2001, \aap, 365, 514

\bibitem[Hillenbrand(1997)]{hil97}
Hillenbrand, L. A. 1997, \aj, 113, 1733

\bibitem[Hughes et al.(1993)]{hug93}
Hughes, J., Hartigan, P., \& Clampitt, L. 1993, \aj, 105, 571

\bibitem[Inaba et al.(2003)]{ina03}
Inaba, S., Wetherill, G. W., \& Ikoma, M. 2003, Icarus, 166, 46

\bibitem[Itoh et al.(2005)]{ito05}
Itoh, Y., et al. 2005, \apj, 620, 984

\bibitem[Janson et al.(2006)]{jan06}
Janson, M., Brandner, W., Henning, T., \& Zinnecker, H. 2006, \aap, 453, 609

\bibitem[Kirkpatrick, Henry, \& Irwin(1997)]{kir97}
Kirkpatrick, J. D., Henry, T. J., \& Irwin, M. J. 1997, \aj, 113, 1421

\bibitem[Kirkpatrick et al.(1999)]{kir99}
Kirkpatrick, J. D., et al. 1999, \apj, 519, 802

\bibitem[Kirkpatrick et al.(2006)]{kir06}
Kirkpatrick, J. D., et al. 2006, \apj, 639, 1120

\bibitem[Klein et al.(2003)]{kle03}
Klein, R., Apai, D., Pascucci, I., Henning, Th., Waters, L. B. F. M. 2003,
\apj, 593, L57

\bibitem[Kraus, White, \& Hillenbrand(2005)]{kra05}
Kraus, A. L., White, R. J., \& Hillenbrand, L. A. 2005, \apj, 633, 452

\bibitem[Kraus, White, \& Hillenbrand(2006)]{kra06}
Kraus, A. L., White, R. J., \& Hillenbrand, L. A. 2006, \apj, in press

\bibitem[Lodato, Delgado-Donate, \& Clarke(2005)]{lod05}
Lodato, G., Delgado-Donate, E., \& Clarke, C. J. 2005, \mnras, 364, L91

\bibitem[Lowrance et al.(1999)]{low99}
Lowrance, P. J., et al. 1999, \apj, 512, L69

\bibitem[Lowrance et al.(2000)]{low00}
Lowrance, P. J., et al. 2000, \apj, 541, 390

\bibitem[Lowrance et al.(2005)]{low05} 
Lowrance, P. J., et al. 2005, \aj, 130, 1845

\bibitem[Lucas et al.(2001)]{luc01}
Lucas, P. W., Roche, P. F., Allard, F., \& Hauschildt, P. H. 2001,
\mnras, 326, 695

\bibitem[Luhman(2004a)]{luh04cha}
Luhman, K. L. 2004a, \apj, 602, 816

\bibitem[Luhman(2004b)]{luh04bin}
Luhman, K. L. 2004b, \apj, 614, 398

\bibitem[Luhman(2006)]{luh06tau}
Luhman, K. L. 2006, \apj, in press

\bibitem[Luhman et al.(2005a)]{luh05wfpc}
Luhman, K. L., McLeod, K. K., \& Goldenson, N. 2005a, \apj, 623, 1141

\bibitem[Luhman et al.(2004)]{luh04ots}
Luhman, K. L., Peterson, D. E., \& Megeath, S. T. 2004, \apj, 617, 565

\bibitem[Luhman \& Potter(2006)]{luh06}
Luhman, K. L., \& Potter, D. 2006, \apj, 638, 887

\bibitem[Luhman et al.(2005b)]{luh05abdor}
Luhman, K. L., Stauffer, J. R., \& Mamajek, E. E. 2005b, \apj, 628, L69

\bibitem[Luhman et al.(2003)]{luh03}
Luhman, K. L., et al. 2003, \apj, 593, 1093

\bibitem[Luhman et al.(2005c)]{luh05cha}
Luhman, K. L., et al. 2005c, \apj, 635, L93

\bibitem[Mamajek(2005)]{mam05} 
Mamajek, E. E. 2005, \apj, 634, 1385

\bibitem[Butler \& Marcy(1996)]{mb96}
Butler, R. P., \& Marcy, G. W. 1996, \apj, 464, L153

\bibitem[Marcy et al.(2005)]{mar05}
Marcy, G., et al. 2005, Progress of Theoretical Physics Supplement, 158, 24

\bibitem[Mart{\'\i}n et al.(1998)]{mar98}
Mart{\'\i}n, E. L., et al. 1998, \apj, 509, L113

\bibitem[Mart{\'\i}n et al.(2000)]{mar00}
Mart{\'\i}n, E. L., et al. 2000, \apj, 543, 299

\bibitem[Mart{\'\i}n et al.(2003)]{mar03}
Mart{\'\i}n, E. L., Barrado y Navascu\'{e}s, D., Baraffe, I., Bouy, H.,
\& Dahm, S. 2003, \apj, 594, 525

\bibitem[Mayor \& Queloz(1995)]{mq95}
Mayor, M., \& Queloz, D. 1995, Nature, 378, 355

\bibitem[McCarthy \& Zuckerman(2004)]{mz04}
McCarthy, C., \& Zuckerman, B. 2004, \aj, 127, 2871

\bibitem[Metchev \& Hillenbrand(2004)]{mh04}
Metchev, S. A., \& Hillenbrand, L. A. 2004, \apj, 617, 1330

\bibitem[Neuh\"{a}user et al.(2002)]{neu02}
Neuh\"{a}user, R., Guenther, E., Mugrauer, M., Ott, T., \& Eckart, A. 2002,
\aap, 395, 877

\bibitem[Neuh\"{a}user et al.(2000a)]{neu00a}
Neuh\"{a}user, R., et al. 2000a, \aap, 354, L9

\bibitem[Neuh\"{a}user et al.(2000b)]{neu00b}
Neuh\"{a}user, R., et al. 2000b, \aap, 360, L39

\bibitem[Neuh\"{a}user et al.(2005)]{neu05}
Neuh\"{a}user, R., et al. 2005, \aap, 435, L13

\bibitem[Oppenheimer et al.(1995)]{opp95}
Oppenheimer, B. R., Kulkarni, S. R., Nakajima, T., \& Matthews, K. 1995, 
Science, 270, 1478

\bibitem[Perryman et al.(1997)]{per97}
Perryman, M. A. C., et al. 1997, \aap, 323, L49

\bibitem[Pollack et al.(1996)]{pol96}
Pollack, J. B., et al. 1996, Icarus, 124, 62

\bibitem[Rebolo et al.(1998)]{reb98}
Rebolo, R., et al. 1998, Science, 282, 1309   

\bibitem[Rieke \& Lebofsky(1985)]{rl85}
Rieke, G. H., \& Lebofsky, M. J. 1985, \apj, 288, 618

\bibitem[Saumon et al.(1996)]{sau96}
Saumon, D., et al. 1996, \apj, 460, 993

\bibitem[Scholz et al.(2006)]{sch06}
Scholz, A., Jayawardhana, R., \& Wood, K. 2006, \apj, in press

\bibitem[Stassun et al.(2006)]{sta06}
Stassun, K. G., Mathieu, R. D., \& Valenti, J. A. 2006, Nature, 440, 311

\bibitem[Vacca et al.(2003)]{vac03}
Vacca, W. D., Cushing, M. C., \& Rayner J. T., 2003, \pasp, 115, 389

\bibitem[White et al.(1999)]{whi99}
White, R. J., Ghez, A. M., Reid, I. N., \& Schultz, G. 1999, \apj, 520, 811

\bibitem[Whittet et al.(1997)]{whi97}
Whittet, D. C. B., et al. 1997, \aap, 327, 1194

\bibitem[Wichmann et al.(1998)]{wic98}
Wichmann, R., Bastian, U., Krautter, J., Jankovics, I., \& Ruci\'nski, S. M.
1998, \mnras, 301, L39

\bibitem[Wilson et al.(2001)]{wil01}
Wilson, J. C., et al. 2001, \pasp, 113, 227

\bibitem[Wolszczan \& Frail(1992)]{wol92}
Wolszczan, A. \& Frail, D. A. 1992, \nat, 335, 145

\end{thebibliography}
\end{document}